Title

# Rotational effect as the possible cause of the east-west asymmetric crater rims on Ryugu observed by LIDAR data

Authors

Naoyuki Hirata [a,*], Noriyuki Namiki [b,c], Fumi Yoshida [d], Koji Matsumoto [e,c], Hirotomo Noda [e,c], Hiroki Senshu [d], Takahide Mizuno [f], Fuyuto Terui [f], Yoshiaki Ishihara [g, h], Ryuhei Yamada [i], Keiko Yamamoto [b], Shinsuke Abe [j], Rina Noguchi [f], Naru Hirata [i], Yuichi Tsuda [f], Sei-ichiro Watanabe [k,f]

* Corresponding Author E-mail address: hirata@tiger.kobe-u.ac.jp

## Authors' affiliation

[a] Graduate School of Science, Kobe University, Kobe, Japan.
[b] National Astronomical Observatory of Japan, Tokyo, Japan.
[c] The Graduate University for Advanced Studies, SOKENDAI, Kanagawa, Japan.
[d] Chiba Institute of Technology, Chiba, Japan.
[e] National Astronomical Observatory of Japan, Iwate, Japan.
[f] Institute of Space and Astronautical Science, JAXA, Sagamihara, Japan.
[g] National Institute for Environmental Studies, Tsukuba, Ibaraki, Japan.
[h] JAXA Space Exploration Center (JSEC), JAXA, Sagamihara, Japan.
[i] The University of Aizu, Aizu-Wakamatsu, Fukushima, Japan.
[j] Nihon University, Funabashi, Chiba, Japan.
[k] Graduate School of Environmental Studies, Nagoya University, Nagoya, Japan.

**Proposed Running Head:** East–west asymmetry of craters on Ryugu
Editorial Correspondence to:
Dr. Naoyuki Hirata
Kobe University, Rokkodai 1-1 657-0013
Tel/Fax +81-7-8803-6566

Key Words
Asteroid
Impact processes
Geological processes

Highlights
- Major craters on Ryugu have east–west asymmetric rims.
- Effect of rotation assessed as the possible cause of the east‐west asymmetry.
- The effect creates the E-W asymmetry only if rotation period was less than 3.6 h.

Abstract

Asteroid 162173 Ryugu is a rubble-pile asteroid, whose top-shape is compatible with models of deformation by spin up. Rims of major craters on Ryugu have an east–west asymmetric profile; their western crater rims are sharp and tall, while their eastern crater rims are rounded and low. Although there are various possible explanations, we theoretically assess the effect of asteroid rotation as the possible reason for this east–west asymmetry. It is known that the trajectories and fates of ejecta are affected by the rotation. The Coriolis force and the inertial speed of the rotating surface are the factors altering the ejecta trajectories. Consequently, we found that the east–west asymmetric crater rims might be formed as a result of rotation, when the inertial speed of the rotating surface is nearly equal to the first cosmic velocity of the body. In other words, it is possible that the observed east‐west asymmetric rims of the Urashima, Cendrillon, and Kolobok craters were formed when Ryugu's rotation period was ~3.6 h.

# 1. Introduction

Asteroid Ryugu is a top-shaped body, with a mean radius of 448.2 m and an equatorial ridge with a mean altitude of 502.5 m, called Ryujin Dorsum. Ryugu is a rubble pile of loose rocks, likely formed into a spinning-top shape during a period of rapid spin (Watanabe et al., 2019). Although the current rotation period of Ryugu is 7.627 h, the period was 3.5 to 5 h based on the geopotential and slope analysis of Ryujin Dorsum (Watanabe et al., 2019). The rotation period of small asteroids, such as 1999KW4 or Ryugu, might vary with time as a result of the Yarkovsky–O'Keefe–Radzievskii–Paddack (YORP) effect (Walsh et al., 2008). Numerous craters on Ryujin Dorsum suggest that the ridge might be a fossil structure (Hirata et al., 2020).

Ryugu has two distinct geologic regions: the cratered terrain (the region between 300°E-30°E and 40°S-40°N) and the western bulge (160°E – 290°E). The former has twice higher crater density than the global average, while the latter has half lower crater density than the global average (Hirata et al., 2020). Also, the latter has a high albedo, a low number density of large boulders, is topographically high, and has a bluish color as compared to the rest of Ryugu (Sugita et al. 2019). Therefore, it is considered that the cratered terrain is geologically old while the western bulge is relatively young (Hirata et al. 2020). There are 7 named craters on Ryugu; Urashima, Cendrillon, Kolobok, Momotaro, Kintaro, Brabo, and Kibidango (Table 1). The majority of the craters are in the cratered terrain, while only Brabo is in the western bulge. Urashima, Kolobok, Kintaro, and Brabo are located across the equatorial ridge of Ryugu, while Cendrillon, Momotaro, and Kibidango are not.

The light detection and ranging laser altimeter (LIDAR) onboard the Hayabusa 2 probe has measured the topographic profiles of Ryugu (Matsumoto et al., 2020). LIDAR science team reported that Urashima and Kolobok craters show east–west asymmetric profiles; the western crater rims are sharp and tall, while the eastern crater rims are rounded and low; while Brabo crater does not show clear east–west asymmetry (Namiki et al., 2019). Noguchi et al. (2020) shows that Cendrillon crater has an east–west asymmetric profile similar to the profiles of Urashima and Kolobok. Here we show the east–west asymmetric profiles of the 7 major craters (Figs. 1 and 2); Kintaro also shows similar east-west asymmetric profiles, while

Momotaro and Kibidango do not show clear asymmetry. The difference in altitude between the western and eastern rims is roughly 10 meter for Urashima, Kolobok, and Kintaro and 30 meter for Cendrillon. We note that Namiki et al. (in preparation) would report observational finding of the geological contexts, wall slopes, depths, and other asymmetries of craters on Ryugu.

Namiki et al. (2019) proposed that there were several possible explanations for the formation of asymmetric profile: first, an oblique impact from the west; second, the effect of rotation; third, the dust that preferentially accumulates on western inner walls due to dust levitation; fourth, the original undulation existing before cratering. The main purpose of this study is to theoretically assess the effect of asteroid rotation as a possible cause of this east–west asymmetry. Assuming various rotation periods for Ryugu, we numerically calculated the spread of the ejecta blankets and east–west profiles near craters. Our simulation does not assess the other hypotheses described by Namiki et al. (2019).

**1.1. The Coriolis force and the inertial speed of the rotating surface**

The Coriolis force and the inertial speed of the rotating surface are the key factors contributing to the asymmetric ejecta patterns (Dobrovolskis and Burns, 1980; Davis et al., 1981). First, due to the Coriolis force, a particle moving inward to the rotational axis is deflected to the direction of rotation in the body-fixed coordinate system (i.e., move to the east), while a particle moving away from the rotational axis is deflected to the direction opposite of rotation (i.e., move to the west). For example, in the case of ejecta launched from the equator, all the ejecta are launched away from the rotational axis. Consequently, their trajectories are deflected to the west, and the ejecta should preferentially accumulate on the western side of the crater. Second, a particle launched to the east has an advantage of the inertial speed of the rotating surface; consequently, the escape velocity for such particles is lower than those launched from the west. If the sum of the inertial speed of the rotating surface (i.e. $2\pi R_a/T$) and the initial launch velocity of a particle relative to the surface overcomes the first cosmic velocity of the body (i.e. $\sqrt{GM/R_a}$), the particle escapes from the body (Dobrovolskis and Burns, 1980; Davis et al., 1981). In the case of a particle launched from the equator, the escape velocity ($v_{es}$) of the particle is written as

$$v_{es} = \sqrt{\frac{GM}{R_a}} - \frac{2\pi R_a}{T} \quad (1)$$

for a particle launched to the east and

$$v_{es} = \sqrt{\frac{GM}{R_a}} + \frac{2\pi R_a}{T} \quad (2)$$

for a particle launched to the west. It is clear that Eq. (1) < Eq. (2), in other words, ejecta thrown towards the east escape at lower launch velocities compared to those thrown towards the west. Here $G$ is the gravity constant and $M, T, R_a$ are the mass, rotational period, and equatorial radius of the body, respectively. In Eqs. (1) and (2), the first and second terms on the right side quantify the first cosmic velocity and the inertial speed of the rotating surface, respectively. When inertial speed of the rotating surface is equal to the first cosmic velocity (i.e. $\sqrt{GM/R_a} = 2\pi R_a/T$), the rotational period is written as

$$T = 2\pi \sqrt{\frac{R_a^3}{GM}}. \quad (3)$$

When the rotation period is shorter than Eq. (3), the escape velocity of a particle launched to the east becomes less than zero, $v_{es} < 0$, and then, all ejecta launched from the equator and thrown towards the east escape from the body and the ejecta blankets should not appear in the east. Assuming the mass and equatorial radius of Ryugu, $M = 4.5 \times 10^{11}$ kg and $R_a = 502.5$ m (Watanabe et al., 2019), such a critical rotation period is $T = 3.59$ h from Eq. (3). Assuming the mass and mean radius of Ryugu, $R_a = 448.2$ m, it is 3.02 h. If the asteroid has a spherical shape and homogenous interior using the density of asteroid ($\rho$), Eq. (3) can be written as

$$T = \sqrt{\frac{3\pi}{G\rho}} \quad (4).$$

Therefore, the rotation period shown in Eq. (3) is basically independent of the asteroid radius. Eq. (4) is known as the critical limit of the rotational period, which is derived by equating the acceleration of gravity at the surface with the centrifugal acceleration at the equator (Pravec and Harris, 2000). Rubble pile bodies aggregated with negligible tensile strength would be disrupted by the centrifugal forces, if the rotational period is less than the critical limit given by Eq. (4).

## 1.2. Previous works for rotational effects and fates of ejecta.

Dobrovolskis and Burns (1980) have examined the fate of crater ejecta from martian satellites by numerical integration of particle trajectories initialized from the satellite surfaces. They showed that the landing locations of ejecta are elongated towards the west because of satellite rotation through the Coriolis effect. They also demonstrated that ejecta thrown towards the east escape at lower launch velocities compared to those thrown towards the west, as indicated in Eqs. (1) and (2), when the rotation period is close to the critical rotation period indicated in Eq. (3). Finally, the trajectories calculated included many interesting features, such as loops, cusps, and folded ejecta blankets. The current numerical analysis for the case of Ryugu is similar to the work of Dobrovolskis and Burns (1980) for Phobos.

The rotation rate of Phobos is gradually accelerating as Phobos approaches Mars as a result of the tidal effect (Shor, 1975; Burns, 1978). Consequently, the past rotational period of Phobos was longer than the current rotational period. For this reason, the trajectories of ejecta from older craters should have been less affected by rotation. Davis et al. (1981) have proposed that assuming the past rotational period to be slightly longer than the current value, observed locations of the so-called grooves correspond to the landing locations of the Stickney-crater's ejecta. Similarly, Thomas (1998) utilized the same assumption regarding the temporal dependence of the rotational period and has proposed that ejecta blankets from the Stickney crater correspond to the so-called blue unit of Phobos's surface if the past rotational period of Phobos was slightly longer.

Asteroid 243 Ida is an irregular, highly elongated asteroid with a rapid rotational period of 4.63 h (Binzel et al., 1993). Geissler et al. (1996) proposed that ejecta blankets have varying morphology between the leading rotational surface (i.e., the area where the zenith roughly corresponds to the rotational direction) and trailing rotational surface (i.e., the area where the zenith roughly corresponds to the direction opposite of the asteroid rotation). For example, ejecta launched from the leading rotational surface escape even at very low launch velocities because of the inertial speed of the rotating surface. In contrast, ejecta launched from trailing rotational surface rarely escape. Moreover, ejecta launched from a given location tend to preferentially accumulate on the leading rotational surface. For example, Geissler et al. (1996) showed that, taking rotation into account, the so-called bluer spectral units on Ida (Veverka et al., 1996) result from Azuura crater ejecta. Asteroid 433 Eros is also a highly elongated asteroid with a rapid rotational period of 5.27 h. Thomas et al. (2001) took the effect of rotation and the irregular shape of the asteroid Eros into account while calculating trajectories of ejecta from a few major craters. Based on the resulting trajectories, it was proposed that most large ejecta blocks on Eros originate from a relatively young crater, Charlois Regio.

## 2. Analysis
### 2.1. Initial launch position, velocity, and volume

The initial launch position, velocity, and volume of ejecta launched from a crater are based upon the scaling law developed by Housen and Holsapple, (2011).

In general, initial launch velocity and volume of ejecta are axially symmetric and depend upon the distance from the center of the crater; ejecta velocities decrease with increasing distance from the crater center. Therefore, ejecta near the crater center achieve large distances or escape velocity, while ejecta near the crater rim accumulate locally. Although the motion of a particle in the excavation flow might be affected by the Coriolis force, we assume it is negligible.

Based on Housen and Holsapple, (2011), the initial launch velocity ($v_{ej}$) of a particle in gravity regime is written as:

$$v_{ej} = C_1 \left( H_1 \sqrt[3]{\frac{4\pi}{3}} \right)^{-\frac{2+\mu}{2\mu}} \sqrt{gR_c} \left( \frac{x}{R_c} \right)^{-\frac{1}{\mu}} \left( 1 - \frac{x}{n_2 R_c} \right)^p , \quad (n_1 a \leq x \leq n_2 R_c)$$
(5)

where $x$ is the distance between the launch position and impact point, $g$ is the surface gravity, and $R_c$ is the apparent crater radius, and $a$ is the projectile radius. Also, the initial launch velocity ($v_{ej}$) of a particle in strength regime is written as:

$$v_{ej} = C_1 \left( H_2 \sqrt[3]{\frac{4\pi}{3}} \right)^{-\frac{1}{\mu}} \sqrt{\frac{Y}{\rho_t}} \left( \frac{x}{R_c} \right)^{-\frac{1}{\mu}} \left( 1 - \frac{x}{n_2 R_c} \right)^p , \quad (n_1 a \leq x \leq n_2 R_c) \quad (6)$$

where $\rho_t$ and $Y$ are the target density and strength. The remaining constants $(\mu, C_1, H_1, H_2, n_1, n_2, p)$ are scaling parameters determined by each ejecta model (Table 2); Housen and Holsapple give 8 sets (C1-C8) of scaling parameters for a variety of materials. In Section 3, we basically used the scaling parameters of C4: dry sand targets in gravity regime, because we consider that gravity regime cratering is more plausible for Ryugu's surface than the strength regime, because Ryugu is a rubble-pile body covered by a thick regolith layer made of non-cohesive material such as sands and boulders (Sugita et al., 2019). This view is supported by the artificial crater formed by the Small Carry-on Impactor (SCI) instrument onboard the Hayabusa 2 probe, which suggested that the crater was formed in the gravity-dominated regime (Arakawa et al., 2020). Nonetheless, the material properties of Ryugu are poorly known and the large craters analyzed in this paper far exceed the size of the artificial crater. Thus, in order to evaluate the difference in sets of scaling parameters, we made Section 3.2. The point at which a particle crosses through the original surface defines the initial launch position ($x$) and velocity ($v_{ej}$) of the particle (Fig. 3). Here, $x$ and $R_c$ are defined as the great-circle distance on a sphere. The launch angle of ejecta is 45° from the target surface in any $x$ (Housen and Holsapple, 2011). We assume $n_1 a = 0$, because the projectile size is much smaller than the crater size. Note that $R_c$ is not the crater rim radius (i.e., the distance between the rim crest and the crater center) but the apparent crater radius (i.e., the distance between the crater center and the position at which a crater inner wall crosses the original surface) as shown in Fig. 3. Thus, $n_2 R_c$ represents the crater rim radius. It should be noted that most of the impact crater database, including Hirata et al. (2020), utilizes crater rim radius or diameter.

The total volume ejected from inside $x$ (i.e., ejecta volume with an initial velocity greater than $v_{ej}(x)$) is given by (Housen and Holsapple, 2011)
$$V = kx^3, \quad (0 \leq x \leq n_2 R_c) \quad (7)$$
where $k$ is a constant determined by each ejecta model (Table 2). To discretize Eq. (7), we assume the ejecta volume at launch is axially symmetric. The volume of a particle launched at $(x_i, \theta_j)$ in polar coordinates is represented by the ejecta volume launched from a small area bounded by a radial distance between $x_{i+1}$ and $x_i$ and a small angular range of $\Delta\theta$:

$$V_{i,j} = k \frac{\Delta\theta}{2\pi}[x_i^3 - (x_i - \Delta x)^3] \ , (8)$$

where

$$x_i = i\Delta x, \ \Delta x = \frac{n_2 R_c}{N_1}, \Delta\theta = \frac{2\pi}{N_2}, i = \{1, \ldots, N_1\}, j = \{1, \ldots, N_2\} \ . (9)$$

In this study, we set $N_1 = 10{,}000, N_2 = 3600$.

## 2.2. Trajectories of ejecta.

For the analysis of ejecta motion close to the asteroid surface, it is more convenient to transform the equations onto an asteroid-fixed frame (Scheeres et al., 1996). When the asteroid rotates with an angular velocity given by the vector $\mathbf{\Omega} = \{0, 0, \omega\}$ in the inertial reference frame, the equation of motion in the asteroid-fixed frame is given by (Scheeres et al., 1996)

$$\ddot{\mathbf{r}} + 2\mathbf{\Omega} \times \dot{\mathbf{r}} + \mathbf{\Omega} \times \mathbf{\Omega} \times \mathbf{r} = -\frac{GM}{|\mathbf{r}|^3}\mathbf{r} \ , (10)$$

where $\mathbf{r}$ is the position vector of a particle relative to the asteroid center and $M$ is the mass of the asteroid. The origin of the coordinate system is the center of the asteroid, the z-axis is taken as the rotational axis, and the xy−plane is taken as the equatorial plane of Ryugu. We assume the shape of Ryugu to be a sphere with a radius of $R_a =448.2$ m. The rotation rate is written as $\omega = 2\pi/(T \times 3600)$ rad/s, where $T$ is given in hours. In this paper, we set $T = 3.0, 3.5, 5.0, 7.627$, and $10{,}000$ h; $T = 10{,}000$ h signifies that the effects of asteroid rotation are negligible, $T = 7.627$ h being the current rotation period of Ryugu, and $T = 3.0$ h is the critical rotation period indicated in Eq. (3).

Given the initial launch position and velocity of the particle described by Eq. (5) or (6), the trajectory of the particle was obtained by numerically integrating Eq. (10) until one of the three outcomes occurs: first, the particle reached an altitude greater than three asteroid radii (i.e., $|\mathbf{r}| > 3R_a$), second, the particle was below the asteroid surface ($|\mathbf{r}| < R_a$), lastly, the particle revolved around the asteroid more than five times. The trajectories are calculated in 1 s steps. When the particle location is given by $|\mathbf{r}| < R_a$, or in other words, when the particle lands on the asteroid surface, the location (latitude and longitude) of the particle is recorded. Ejecta thickness is given by dividing the sum of the volume of particles falling within a 1° colatitude radius circle around the point by the area of the 1° colatitude circle. We do not consider lateral movement after landing, such as subsequent downslope motion or secondary ejection. Therefore, the ejecta thickness at crater rim (i.e. $n_2 R_c$ from the crater center) could result in unrealistically steep slopes. The actual ejecta thickness should be slightly more diffused than this estimate. Although it is not satisfactory for determining the maximum height of the crater rim, it is considered suitable for obtaining the approximate volume or the presence of ejecta rim.

We assume the shape of the asteroid is the sphere and ignored the topographical effects on Ryugu and complex gravity fields. Although topographical effects result in shielding due to mounds around craters on Phobos and Eros (Thomas 1998; Thomas et al., 2001), the sphere is more appropriate assumption for this type of study, because if we use the actual shape of Ryugu, which has already appeared the east-west asymmetric profiles, we cannot judge whether our model results is attributed to the effect of original topography or the effect of rotation.

We note that our analysis had the following limitations: 1) we assumed a launch angle (measured from the surface) of 45° based on Housen and Holsapple (2011); however, it was not clear if this assumption was suitable (for example, Thomas (1998) assumed 32°) and 2) crater rims are composed of ejecta deposition and syncline, but we only assessed the ejecta deposition.

## 3. Results
### 3. 1. Case of the ejecta from a crater with a diameter of 100 m

We defer results for craters on Ryugu taking a simple example: a crater with a rim radius of $n_2 R_c = 50$ m, formed at the equator, 0° N and 180° E.

#### 3.1.1. Trajectories of ejecta particles

Figure 4 and 5 show the trajectories of ejecta launched to the west and east from a 50 m radius crater. When $T = 10,000$ h, the trajectories of ejecta are symmetric (Fig. 4 a, b), while when $T \leq 7.627$ h, the trajectories of ejecta are no longer symmetric (Fig. 4d-g and 5c, d).

Figures 6 and 7 show landing locations of particles launched from the craters as a function of initial launch velocity and direction. Each contour in Fig. 6 signifies landing locations of particles launched with identical initial launch velocities. When $T = 10,000$ h (Fig. 6a), the contours are concentric patterns. When $T$ = 7.627 h (Fig. 6b), the contours of relatively low launch velocity (< 14.9 cm/s) are dense west of the crater and are sparse east of the crater. This is consistent with Fig. 5 of Davis et al. (1981). The contour for a relatively large launch velocity (20.4 cm/s) is distorted along the east–west axis and is similar to a Cardioid. The contour of 29 cm/s was transposed; the east side of the contour is composed of ejecta launched to the west, while the west side of the contour contains ejecta launched to the east. The non-closure of the west side of this contour signifies that the particles escaped in this direction. The trajectories of particles launched to the north or northeast (south or southeast) are deflected to the southwest (northwest) (Fig. 7b). When $T$ = 5.0 h (Fig. 6c), the contours of low launch velocity (< 8.6 cm/s) are dense west of the crater and sparse east of the crater. The contours of high launch velocity (> 14.9 cm/s) distort and become intersecting curves. When $T$ = 3.0 or 3.5 h (Fig. 6d, e), every contour is elongated to the west. Even a particle launched at the velocity of 1.18 cm/s appears west of the crater (Fig. 7d, e).

### 3.1.2. Global distribution of ejecta thickness

Figure 8 shows the global distribution of ejecta thickness. When $T$ = 10,000 h, the ejecta thickness is symmetric. When $T$ = 7.267 h, the ejecta thickness is asymmetric. A bow-shaped front line appears east of the crater, and ejecta thickness west of the line is negligibly thin. The bow-shaped line is the dividing line for landing locations of particles launched to the east and those launched to the west, as shown in Fig. 6b, c. The ejecta thickness at the front line is slightly thick. This front line probably contains folded ejecta blankets predicted by Dobrovolskis and Burns (1980). As $T$ becomes shorter, the distance between the front line and the crater rim is smaller. When $T$ = 3.5 h, the front line is nearly adjacent to the crater rim. When $T$ = 3.0 h, the front line appears in the west of the crater and the crater rim does not appear east of the crater. Most of ejecta launched to the east accumulate around the equator.

### 3.1.3. East–west profile near the crater

Figure 9 shows the west-east profile (i.e. the profile along latitudinal line) of the ejecta thickness near the same craters shown in Fig. 8. Ejecta thickness profiles near the crater rim (Fig. 9a) are not affected much by the rotation rate when $T$ > 3.5 h. When $T$ = 3.5 h, the ejecta thickness in the east of the crater is slightly more massive than the thickness in the west, because the bow-shaped front line is nearly adjacent to the eastern crater rim. When $T = 3.0$ h, ejecta does not accumulate east of the crater, while ejecta thickness in the west does not vary from that in $T = 10{,}000$ h. Figure 9b shows the east–west profile as a function of $T$, which varies from $T = 3.0$–3.5 in increments of 0.1 h. The characteristics of the east crater rim are very sensitive to the rotation period. When $T > 3.2$ h, the ejecta thickness does not show clear east–west asymmetry, while for $T < 3.1$ h, it shows clear east–west asymmetry. This is very close to the critical rotation period indicated in Eq. (3). In other words, an east–west asymmetric profile for the crater rim is generated only if the inertial speed of the rotating surface is nearly equal to the first cosmic velocity of the body. We consider that, because ejecta thickness near crater rim is formed by ejecta with very low launch velocities and short travel times, the effect of Coriolis forces on the trajectories is rather small. When the rotation period is nearly equal to that given in Eq. (3), the eastern crater rim disappears as a result of the inertial velocity of the rotating surface.

Difference in height between the west and east crater rim is roughly 5 to 8 meters (Fig. 9b). Although, as we stated in Section 2.2, we do not consider lateral movement after landing, this value is mostly consistent with the observed differences for Urashima, Kolobok, and Kintaro (10 meter), but is inconsistent with that for Cendrillon (30 meter).

### 3.2. Case of other sets of scaling parameters.

Adopting 8 sets of scaling parameters for a variety of target materials indicated in Table 2, we calculated and compared results for craters on Ryugu taking a simple example: a crater with a rim radius of $n_2 R_c = 50$ m formed at the equator (Fig. 10). The 8 sets include four examples in gravity regime and four in strength regime. Note that C4 is actually identical to C5 and therefore, we did not show the case of C5 in Fig. 10. As a result, we find that, although the height and width of rims are slightly different in each set, an east–west asymmetric crater rim profile can be generated only if the rotation period is very close to Eq. (3) in any sets. Therefore, the difference of scaling parameters barely affects the result in Section 3.1.

### 3.3. Case of craters formed apart from the equator.

In general, the inertial velocity of a rotating body is maximum at the equator and minimum at the poles. In order to study the latitude dependence of east–west asymmetry, we compare a 50 m radius crater formed at equator (Fig. 8e), 30° N, 180° E (Fig. 11a), and 60° N, 180° E (Fig. 11b). Consequently, we found there is a small asymmetry for 30° N and very small asymmetry for 60° N (Fig. 11c). Note that, majority of the major craters on Ryugu, including Urashima, Cendrillon, Kolobok, and Brabo, exist at low latitudes < 30°.

### 3.4. Case of actual craters on Ryugu

We used the location and size of the four actual craters on Ryugu: Urashima, Cendrillon, Kolobok, and Brabo, and simulated the ejecta thickness as a function of the rotation period.

The model results for Urashima, Kolobok, and Brabo (Figs. 12, Fig. 13) are very similar to those for the simple case study example shown in Section 3.1. We again confirm that, when $T$ = 3.0 h, ejecta do not accumulate near the east rim of the crater, while ejecta thickness near the west rim does not differ from that in the $T$ = 10,000 h simulation. The model results for the Cendrillon crater (Fig. 12f–j) are very similar to the model results shown in Fig. 11a and discussed in Section 3.3. Theoretical results estimate that, when $T$ = 3.0 h, the Cendrillon crater would have weak east–west crater rim asymmetry. The Urashima and Cendrillon (Fig. 12) craters show a north–south asymmetry; the equatorial crater rim sides are thinner than the polar sides.

### 4. Discussion

The results demonstrated that east–west asymmetric crater rim profiles could be generated by the effect of rotation when the inertial speed of the rotating surface is nearly equal to the first cosmic velocity of the body. From Eq. (3), such a critical rotation period is 3 hour in the case of a sphere with a radius of 448.2 m, while it is 3.59 hours in the case of the equatorial radius of 502.5 m. Therefore, adapting the ideal condition shown in Section 3.4 for the equatorial radius of Ryugu, the effect of rotation can explain their observed east–west asymmetric profiles only if Ryugu's rotation period were ~3.6 h. On the other hand, the Brabo crater, not displaying east–west asymmetry, suggests that they were possibly formed after Ryugu was despun. The Brabo crater may be relatively younger than the other three craters. This view is consistent with the fact that the Urashima, Cendrillon, and Kolobok craters exist on the old cratered terrain, while the Brabo crater exists on the young western bulge (Hirabayashi et al., 2019; Hirata et al., 2020). However, there is no way to determine the order of formation for these craters; the stratigraphic relationship of these craters is unclear. As well, the Momotaro and Kibidango craters, not displaying east–west asymmetry, and the Kintaro crater, displaying east–west asymmetry, could be explained by the same manner. On the other hand, although difference in height of ejecta rims is consistent with the observed ones of Urashima, Kolobok, and Kintaro (10 m), it is not consistent with one of Cendrillon (30 m). Therefore, the asymmetry of Cendrillon may be derived from the original undulations before its formation.

The ejecta tend to accumulate on the equator when it was formed during rapid rotation period, which may explain the blue unit along Ryugu's equatorial ridge pointed out by Sugita et al. (2019). On the other hand, we cannot identify the bow-shaped front line near the craters. Possible explanations may be that the front lines is not as thick as observed (Figures 8, 9, 11, 12, 13 are colored in log scale and actual ejecta thickness in the front line is very thin) or the lines are too small to be resolved in the plot. If these explanations are false, the origin of east–west asymmetric crater profiles may not be attributed to rotational effects.

This work can be adapted for other asteroids, as shown in Eq, (4). For example, the Fingal crater, with a diameter of 14 km, is a relatively new crater on Ida. It shows an east–west asymmetric profile with a sharp rim on the west and a rounded rim on the east. This crater is considered to be the result of an oblique impact (Sullivan et al., 1996); however, possible explanations for this asymmetry include the effect of asteroid rotation.

## 5. Conclusions

We theoretically assess the possibility of rotation as the cause of east–west asymmetric rims of the Urashima, Cendrillon, and Kolobok craters. We found that this mechanism creates the asymmetric rims only if the inertial speed of the rotating surface is nearly equal to the first cosmic velocity of the body. In the case of asteroid Ryugu's equator, such a rotation period was 3.6 h. Consequently, the inertial speed of the rotating surface, rather than the Coriolis force, appeared to dominate in mechanisms forming asymmetric east–west crater rim profiles. Our results indicated the possibility that the Urashima, Cendrillon, Kintaro, and Kolobok craters were formed when Ryugu's rotation period was ~3.6 h, while the Brabo, Momotaro, and Kibidango craters were formed after Ryugu initiated a slowing in its rotation rate.

## Acknowledgments

We thank all members of Hayabusa2 mission team for their support of the data acquisition. This work was supported by JSPS Grants-in-Aid for Scientific Research Nos. 20K14538 and 20H04614 (Naoyuki Hirata). All of LIDAR topographic data and ONC images obtained by Hayabusa2 and the shape model of Ryugu will be freely available via Data ARchives and Transmission System (DARTS) at ISAS/JAXA at the end of 2020. We appreciate two anonymous reviewers and their helpful comments.

Table 1. The 7 named craters on Ryugu and their basic data from Hirata et al. (2020).

| Name | Lat. | Lon. (°E) | D (m) | CL[*1] |
|---|---|---|---|---|
| Urashima | -7.19 | 92.99 | 290 | I |
| Cendrillon | 28.34 | 353.68 | 224 | II |
| Kolobok | -0.70 | 330.28 | 221 | II |
| Momotaro | -14.83 | 51.20 | 183 | I |
| Kintaro | 0.42 | 157.84 | 154 | II |
| Brabo | 3.24 | 229.95 | 142 | I |
| Kibidango | -31.50 | 47.26 | 131 | I |

[1] Classification shown in Hirata et al.; CL I means circular depression with rim and II circular depression without rim.

Table 2. Scaling parameters used in ejecta model, based on Housen and Holsapple, (2011).

| No.* | C1 | C2 | C3 | C4 | C5 | C6 | C7 | C8 |
|---|---|---|---|---|---|---|---|---|
| Reg.** | G | S | S | G | G | G | S | S |
| $\mu$ | 0.55 | 0.55 | 0.46 | 0.41 | 0.41 | 0.45 | 0.40 | 0.35 |
| $k$ | 0.2 | 0.3 | 0.3 | 0.3 | 0.3 | 0.5 | 0.3 | 0.32 |
| $C_1$ | 1.5 | 1.50 | 0.18 | 0.55 | 0.55 | 1.00 | 0.55 | 0.60 |
| $H_1$ | 0.68 | - | - | 0.59 | 0.59 | 0.8 | - | - |
| $H_2$ | - | 1.1 | 0.38 | - | - | - | 0.40 | 0.81 |
| $n_2$ | 1.5 | 1.0 | 1.0 | 1.3 | 1.3 | 1.3 | 1.0 | 1.0 |
| $p$ | 0.5 | 0.5 | 0.3 | 0.3 | 0.3 | 0.3 | 0.3 | 0.2 |
| $\rho_t$ (kg/m3) | 1000 | 3000 | 2600 | 1600 | 1510 | 1500 | 1500 | 1200 |
| $Y$ (MPa) | - | 30 | 0.45 | - | - | - | $4\times10^{-3}$ | $2\times10^{-3}$ |

* Scaling parameters for Water (C1), Rock (C2), weakly cemented basalt (C3), sand (C4 and C5), glass micro-spheres (C6), sand/fly ash mixture (C7), and perlite/sand mixture (C8).

** The strength regime (S) or gravity regime (G).

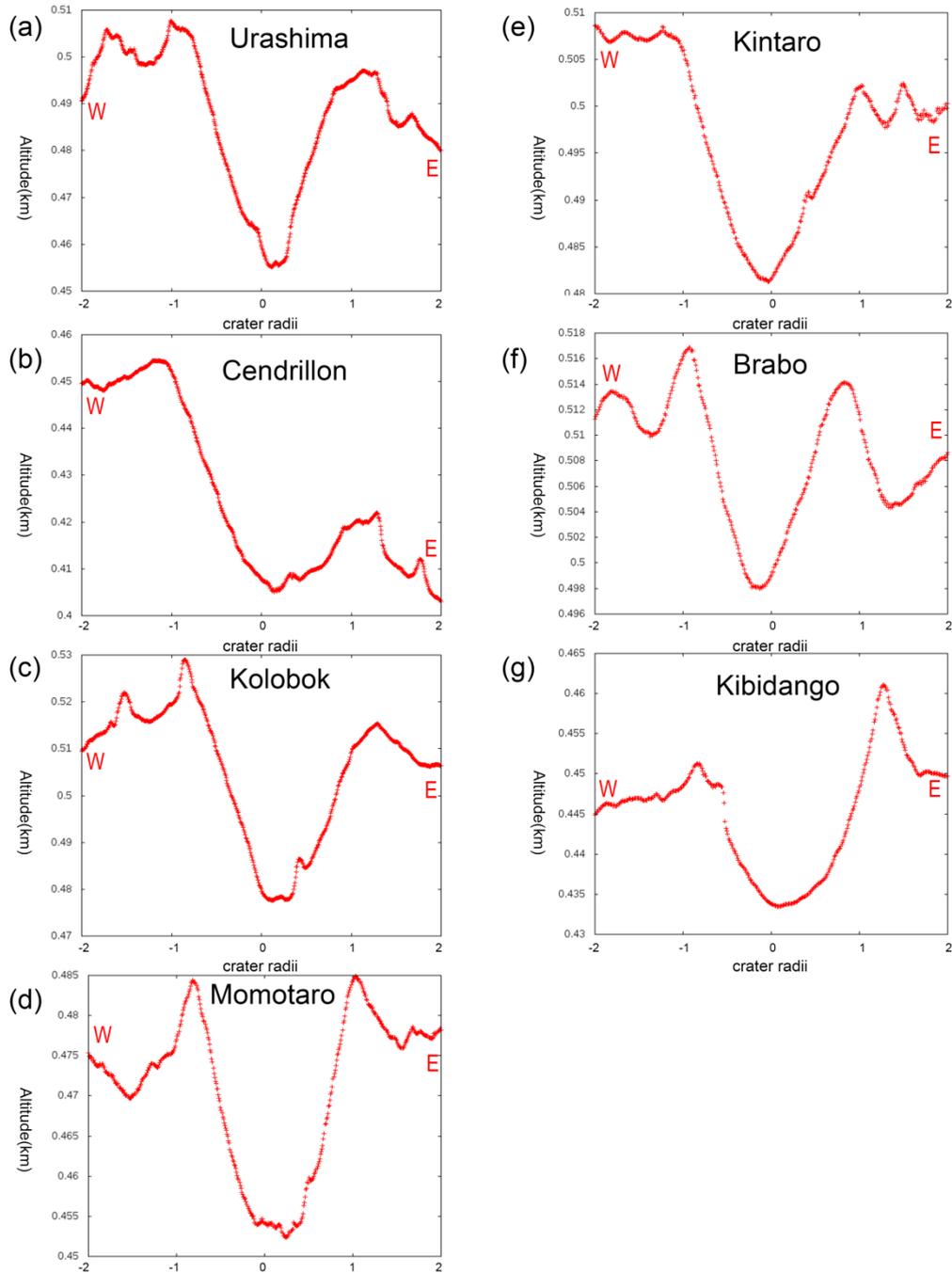

**Figure 1.** Observed east-west asymmetric profiles of the 7 named craters on Ryugu. In each plate, left side is west, and right side is east. The profiles are from the shape model of Watanabe et al. (2019). Survey lines are shown in Figure 2. Note that a sharp peak at the right side of Kibidango, and two sharp peaks at the right side of Kolobok are boulders.

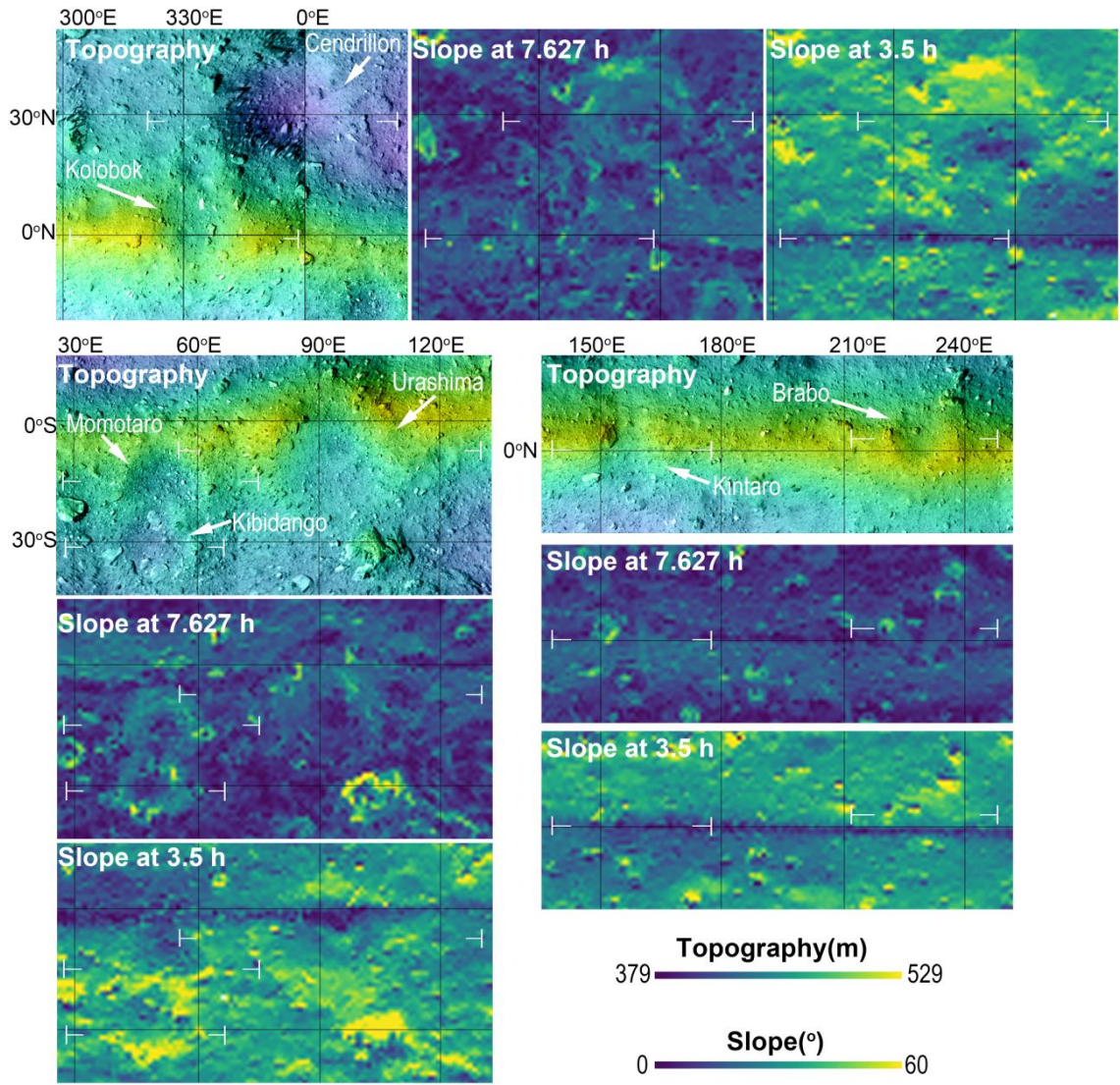

**Figure 2.** Local simple cylindrical projection maps of the 7 named craters and locations of the survey lines in Figure 1. In each location, there are three plates of ONC-T images mosaic colorized by topography and geopotential slope at a rotation of 7.627 h and 3.5 hour. The center and length of the lines is defined as the center and twice of the diameter of each crater, respectively. Topography is defined as the distance from the geometric center of Ryugu.

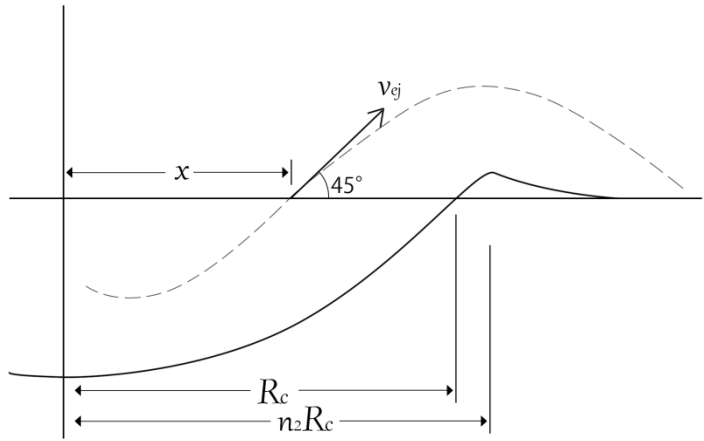

**Figure 3.** Definition of variables in Eq. (5) in Section 2.1. Dashed line is a trajectory of a particle launched at a location, $x$.

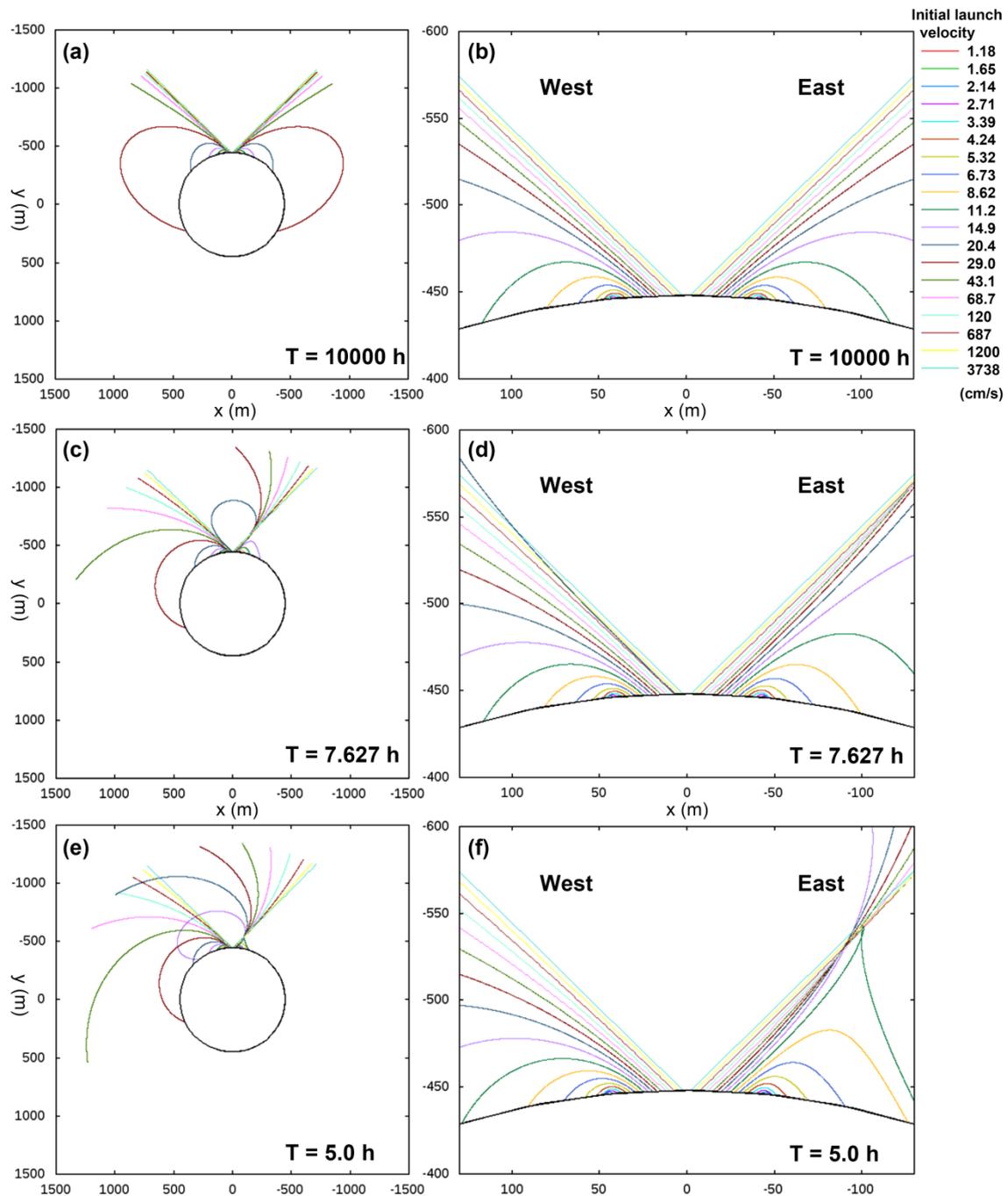

**Figure 4.** The trajectories of a particle launched from a 50m-radius crater to the west and east, when $T$=10000, 7.627, and 5.0 hours. The xy-plane corresponds to the equatorial plane of Ryugu. Although we calculate trajectories of $N_1 \times N_2 = 36{,}000{,}000$ ejecta particles in order to obtain Figs. 7, 9-11, a part of them is shown in this figure. The velocities of the trajectories correspond to the initial lunch velocities of ejecta particles launched from the initial launch position, $x$, in increments of one-twentieth of the crater rim radius.

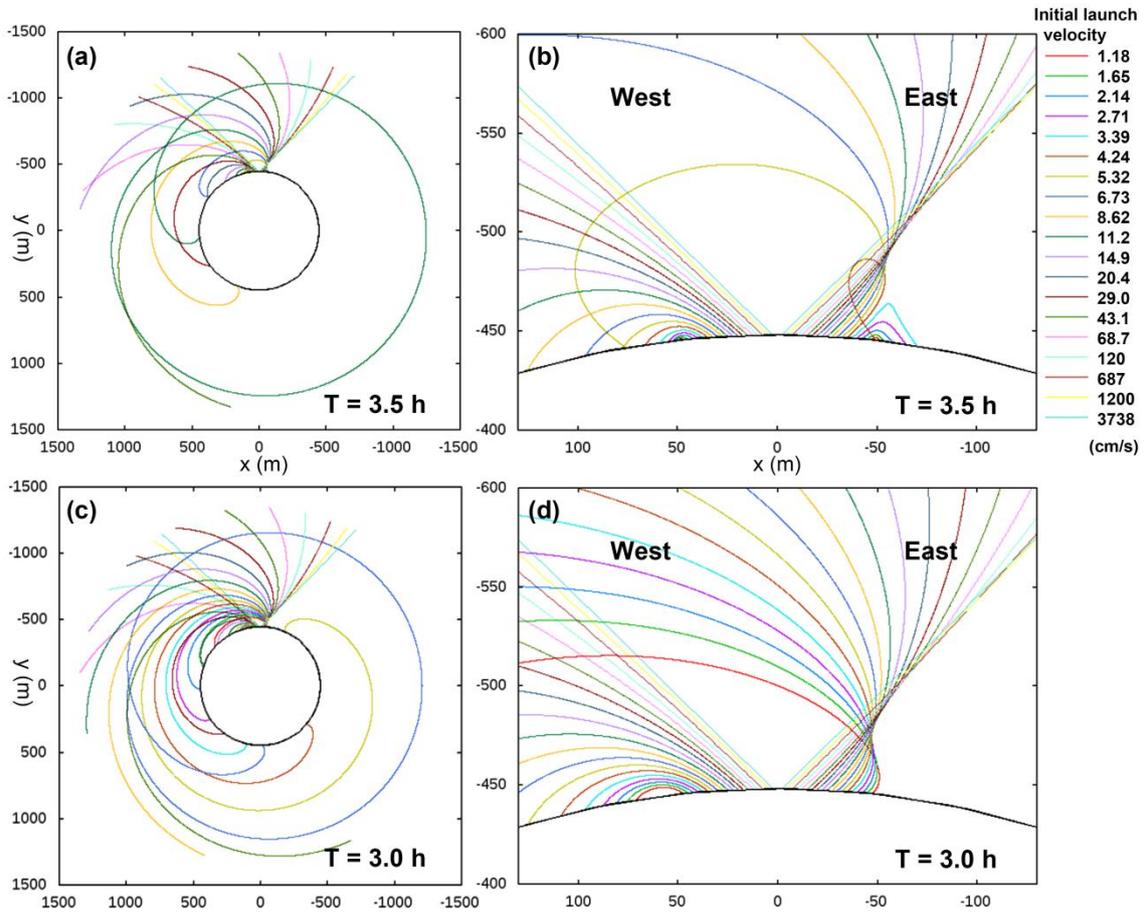

**Figure 5.** The trajectories of a particle launched from the 50m-radius crater to the west and east, when *T*=3.5 and 3.0 hours.

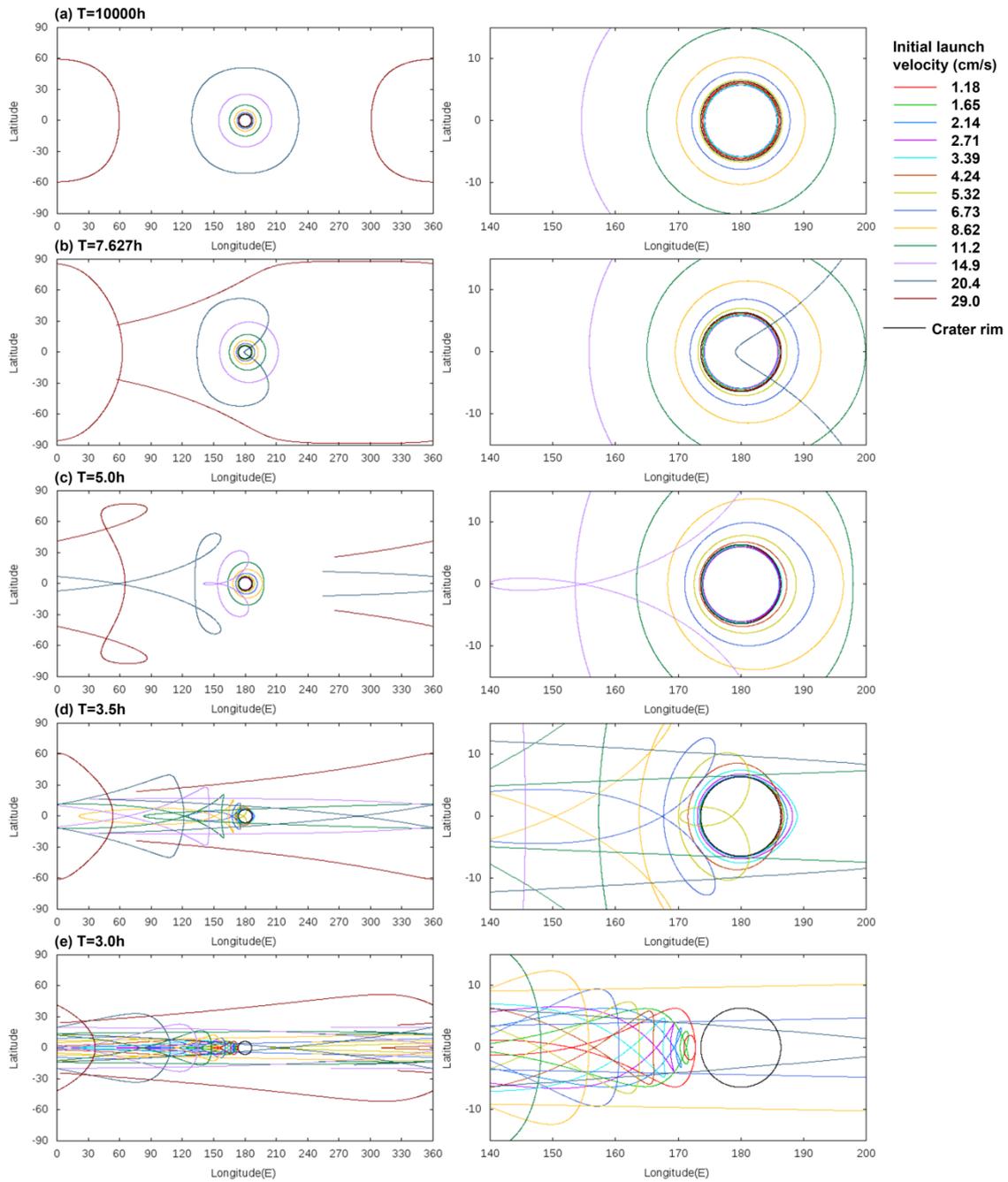

**Figure 6.** The landing locations of particles launched from a crater with a radius of 50m at the equator as a function of initial launch velocity of ejecta particles.

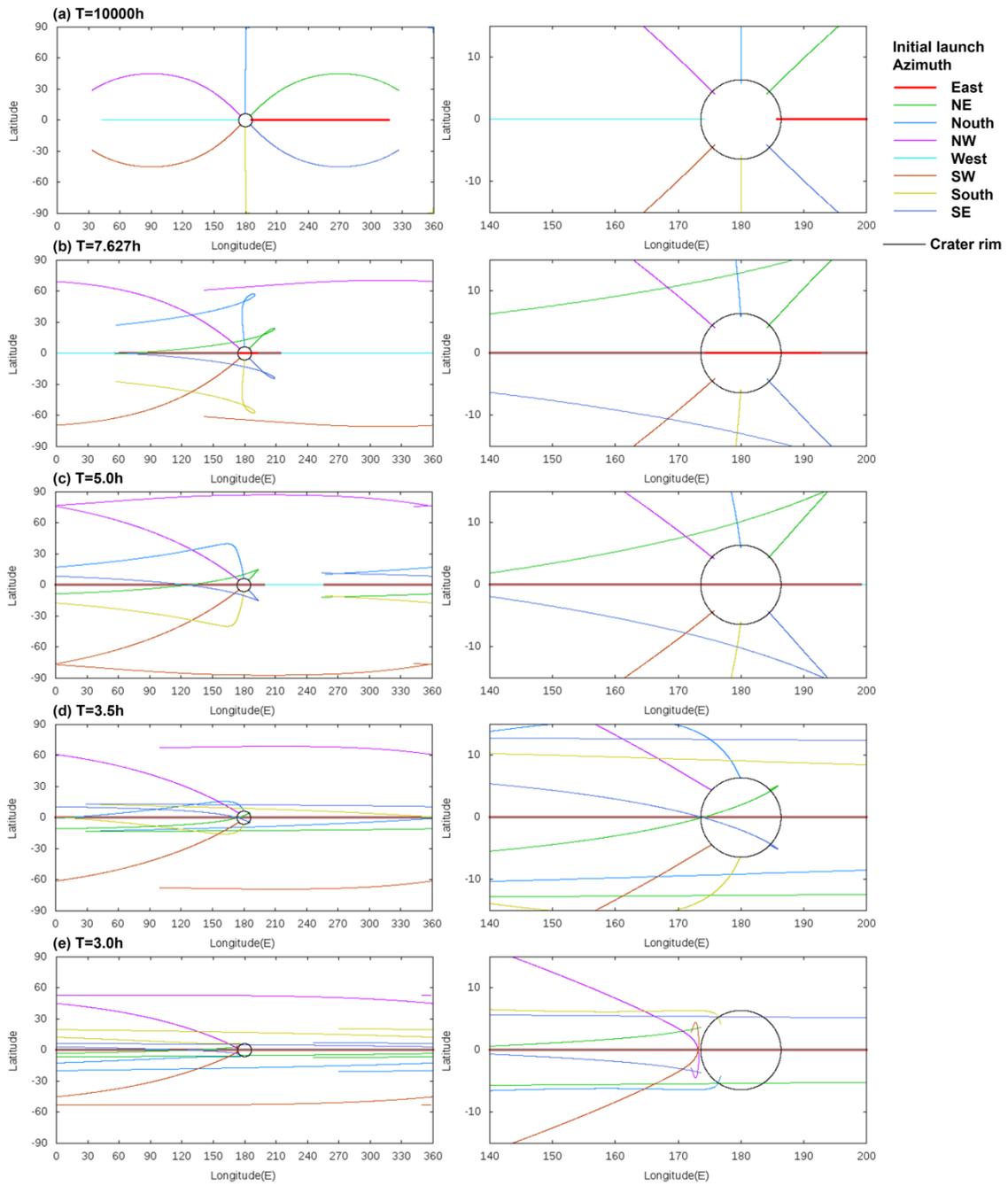

**Figure 7.** The landing locations of particles launched from a crater with a radius of 50m at the equator as a function of initial launch direction of ejecta particles.

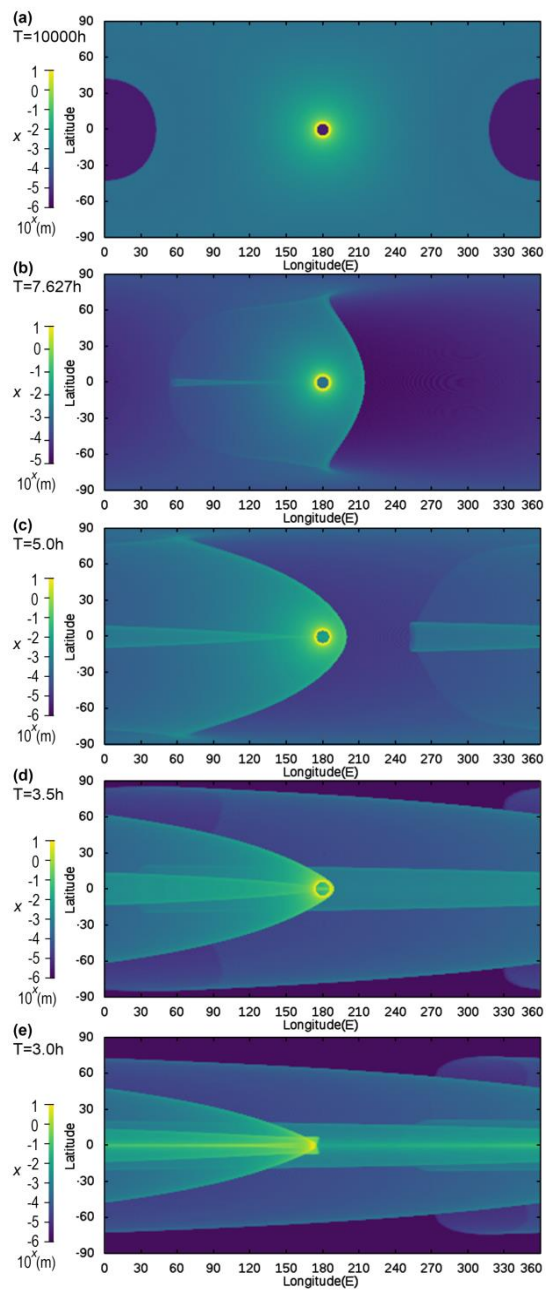

**Figure 8.** The global distribution of ejecta thickness launched from a 50m-radius crater at the equator.

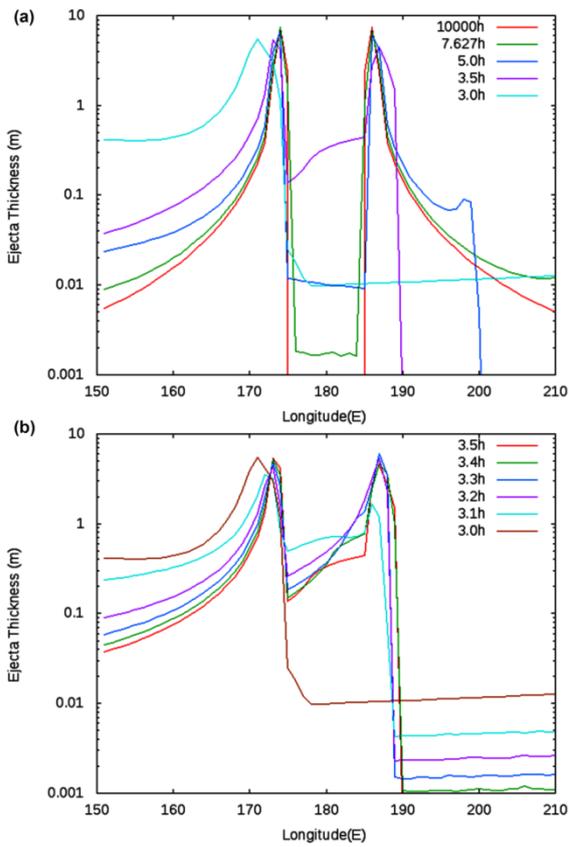

**Figure 9.** The west-east profile of the ejecta thickness near the crater at the equator.

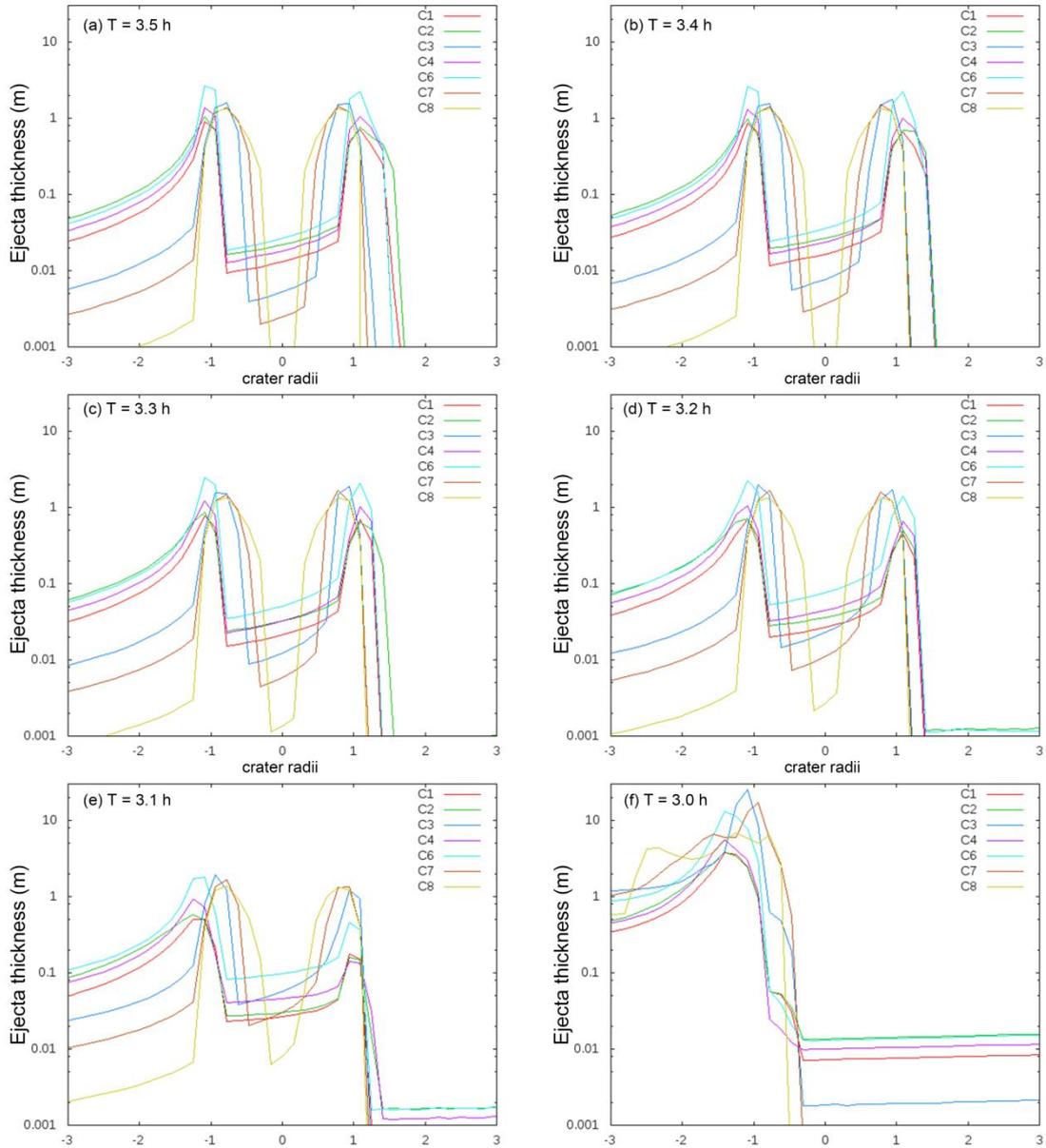

**Figure 10.** The comparison of model results in the 8 sets of scaling parameters in Table 2. It shows that the west-east profile of the ejecta thickness near the crater along the equator as a function of the rotation period between 3.5h and 3.0 h.

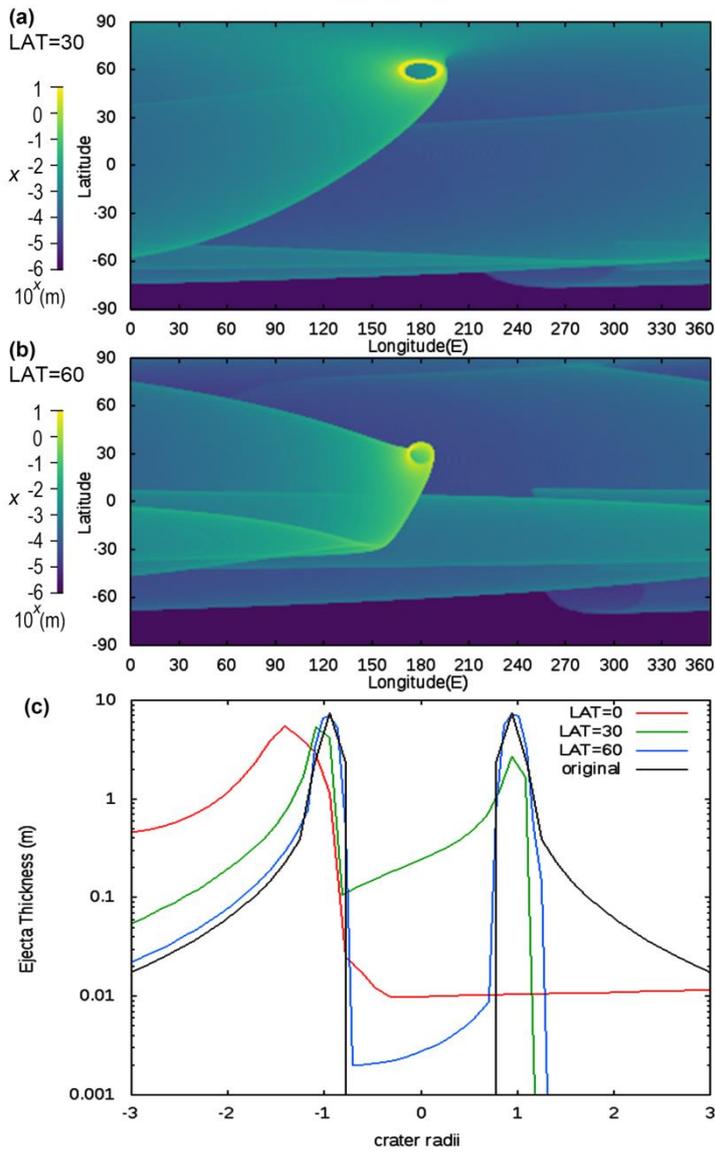

Figure 11. (a) The global ejecta thickness of a 50m-radius crater formed at 30°N, 180°W when $T=3.0$ hours. (b) The global ejecta thickness of a 50m-radius crater formed at 60°N, 180°W when $T=3.0$ hours. (c) The comparison of the west-east profile of the ejecta thickness near the crater formed at the equator, 30°N, and 60°N when $T=3.0$ hours. The horizontal axis is scaled at 1 crater radius (50 meter).

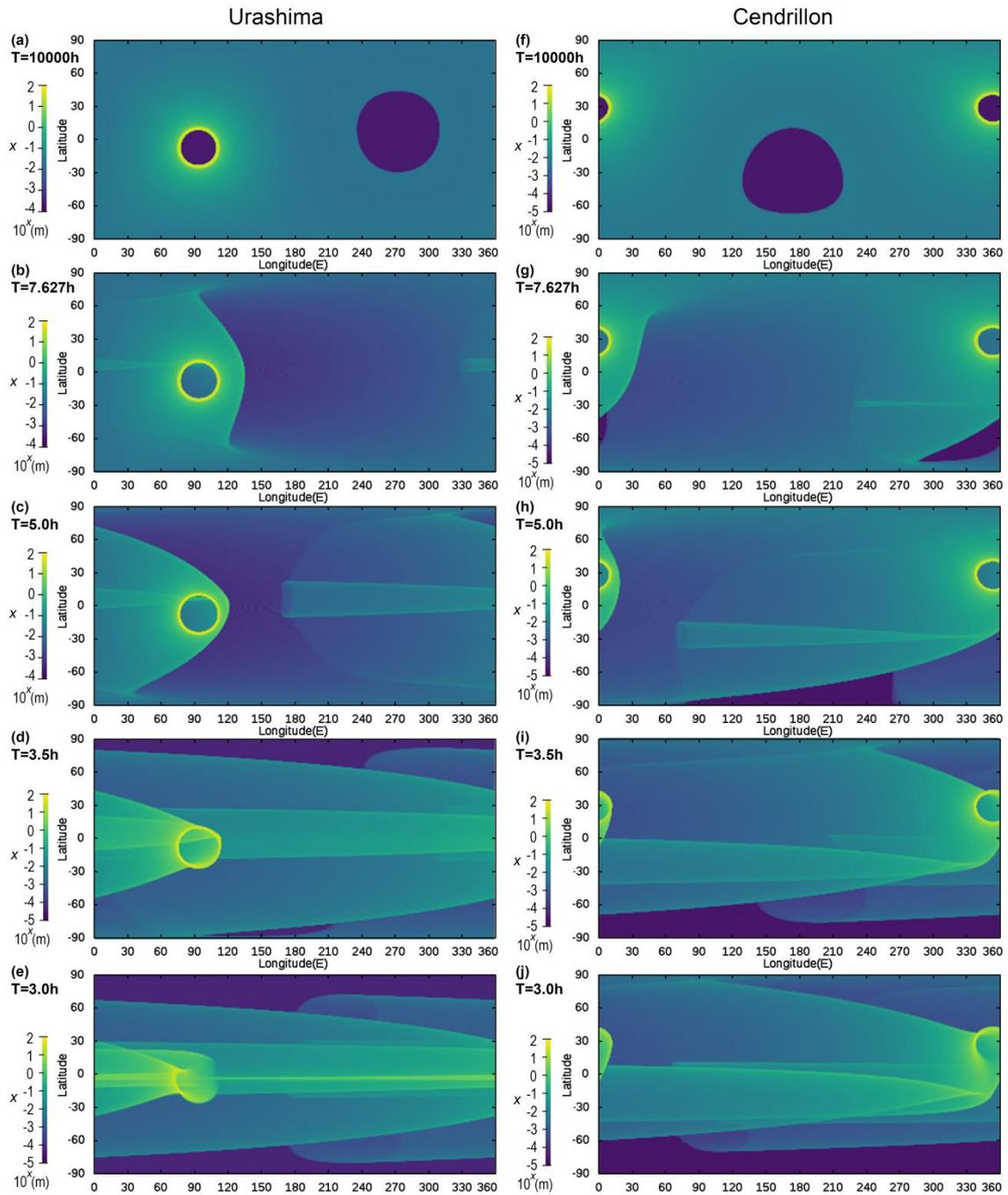

**Figure 12.** The global ejecta thicknesses from Urashima, Cendrillon, and Kolobok and Brabo craters.

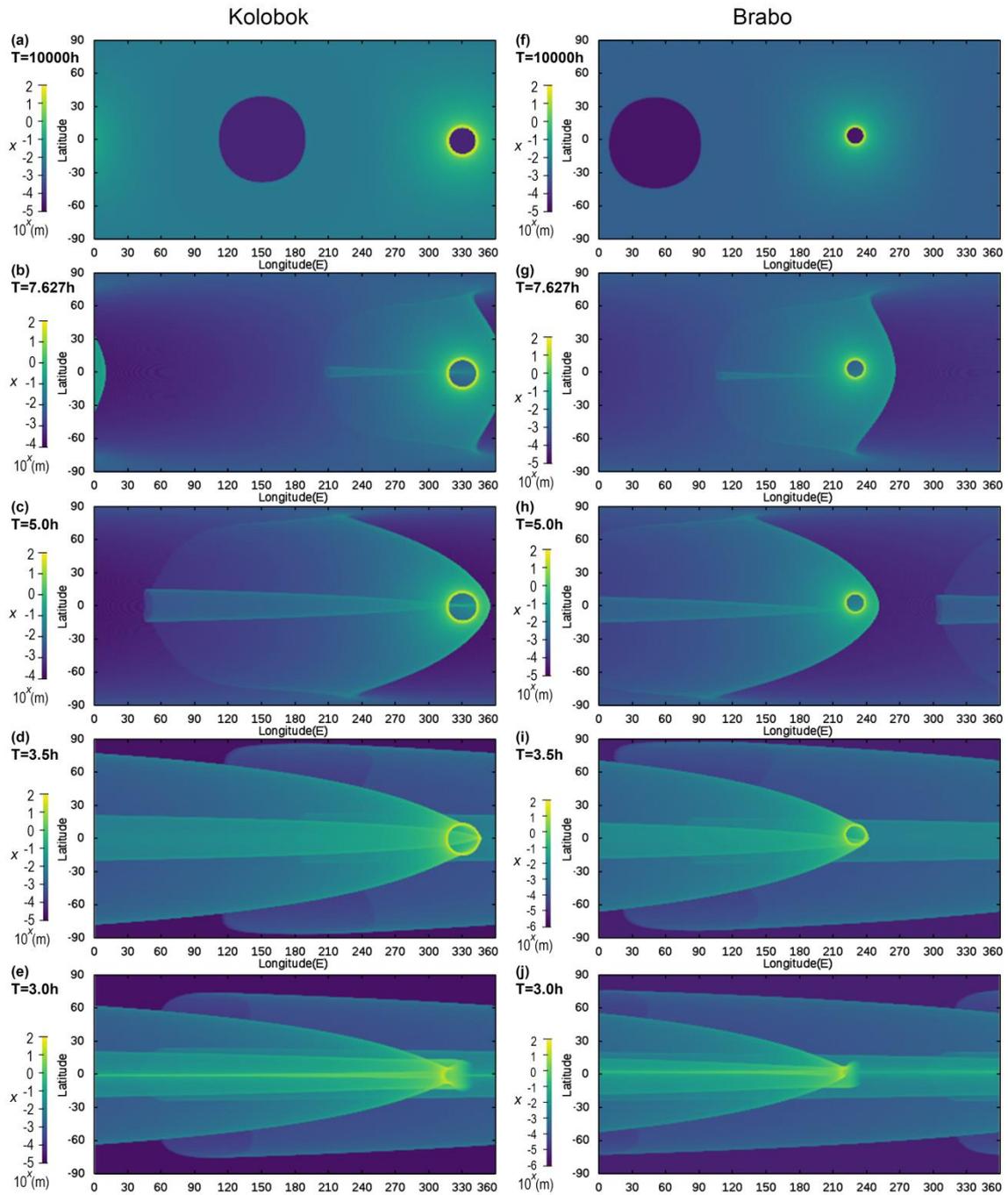

**Figure 13.** The global ejecta thicknesses from Kolobok and Brabo craters.